\documentstyle[12pt]{article}

\begin{document}

\begin{center}
{\Large\bf Poisson-Lie Structures and Quantisation with Constraints}
\end{center}

\begin{center}
Petre Di\c t\u a\footnote{ email: dita@theor1.ifa.ro} 
\end{center}

\begin{center}
{Institute of Theoretical Physics\\ Sidlerstrasse 5, Bern,
Switzerland\
\\and\\ National Institute of
Physics \& Nuclear Engineering\\ P.O.Box MG6, Bucharest, Rumania}

\end{center}

\begin{abstract}
We develop here a simple quantisation formalism  that make use of
Lie algebra properties of the Poisson bracket. When the brackets
$\{H,\varphi_i\}$ and $\{\varphi_i,\varphi_j\}$, where $H$ is the
Hamiltonian and $\varphi_i$ are primary and secondary constraints, can
be expressed as functions of $H$ and $\varphi_i$ themselves, the
Poisson bracket defines a Poisson-Lie structure. When this algebra has a
finite dimension a system of first order partial
differential equations is established whose solutions are the
observables of the theory. The method is illustrated with a few examples.
\end{abstract}
\vskip1cm
1.  The quantisation of systems with constraints is
old as the quantum mechanics itself. The first such problem
brilliantly solved was the finding of the hydrogen atom spectrum by
Pauli in 1926 \cite{Pa}. Enforcing the constraints in classical
mechanics has a satisfactory solution \cite{Wh,MW}, but this is no
more true in quantum mechanics.
The constraints, i.e. a set of functions
$$\varphi_i(q,p)=0~~~,~~~ i=1,2,\dots,m \eqno(1.1)$$
restrict the motion of the classical system to a manifold embedded in the 
initial 
Euclidean phase space and  in consequence  the canonical quantisation rules
$$[q_i,p_j]=i\hbar\delta_{ij}$$
are no more sufficient for the quantum description of the physical system.

In general quantisation is not a well-defined procedure existing today
a variety of methods which sometimes give different results when
applied to physical problems, although the starting points are similar
from the classical point of view. We mention only the people 
who derive the Schr\"odinger equation by Feynman's path integral method; 
see for example
\cite{Wi,Ch,De}, who  find an extra energy term proportional to the 
Riemann scalar curvature
of the manifold, even for the simple case of the motion of a particle
on the $n$-dimensional sphere.

The most succesfull method for imposing quantum constraints is that
found by Dirac \cite{Di}, however nowadays there are some voices who
reject it claiming that the resulting energy spectrum is incorrect
even for simple systems \cite{KS,KlS}. The mechanism found by Dirac was
the introduction of a new symplectic structure, the Dirac bracket, to
handle the second-class constraints and the using of the Legendre
multiplicators for finding the true Hamiltonian. 

The purpose of this paper is to look at the problem of quantisation
with constraints from a slightly modified point of view and to show
that the new proposal leads to correct results.

When one studies  constrained systems one starts with a Hamiltonian, $H(q,p)$,
and a number of relations of the form (1.1), called primary
constraints, which at their turn generate secondary constraints. Let
suppose  that after a finite number of  steps the process closes, i.e.
no new secondary constraints are generated. In the most simple cases
one obtains a Poisson algebra of the form
$$\{H(q,p),\varphi_i(q,p)\}=C_i^j\varphi_j(q,p)\eqno(1.2)$$
$$\{\varphi_i(q,p),\varphi_j(q,p)\}=C_{ij}^k\varphi_k(q,p)$$
where $C_i^j$ and $C_{ij}^k$ are constant structure coefficients.
In our opinion this Poisson structure  is the basic structure for the 
quantisation
procedure. Since the Poisson algebra  (1.2) transforms by quantisation
into a Lie algebra the physical observables of the model will be given
by the Casimir operators; this means that no one of the initial
operators transform into a veritable observable.

We applied this idea to the motion of a particle on the
$n$-dimensional sphere and we have found that the "Hamiltonian",
i.e. the Casimir of the corresponding algebra is a quadratic function
in the old Hamiltonian and the constraints \cite{PD}. This quantity is
the square of the angular momentum, a result which everybody expected
to be so.

We want to extend this method to more general situations than those
given by Eqs. (1.2) by developing a formalism which makes use of the
Lie algebra  properties of the Poisson bracket. We hope that this formalism 
 will
solve  at least a
part
of problems encountered in
quantisation  with constraints.

More precisely let $(u_1,u_2,\dots,u_r)$ denote $r$ functions of $2n$
independent variables $(q_1,\dots,q_n,p_1,\dots,p_n)$ and suppose that
all Poisson brackets $\{u_i,u_j\}$ can be expressed as functions of 
 $(u_1,u_2,\dots,u_r)$. In this case these functions  form a Poisson-Lie
structure and any function of   $(u_1,u_2,\dots,u_r)$
belongs to this algebra.
This kind of structure was first introduced by S. Lie who use the name
of
function group \cite{SL} . The full phase space is $R^{2n}$ with generic
point $({\bf q,p})$ and the usual Poisson algebra ${\cal
P}=(C^{\infty}(R^{2r}),\{\cdot,\cdot\})$ is the setting for the problem.

The systems with constraints are good
candidates for such structures since we start with a Hamiltonian and a
number of primary constraints of the form (1.1). By taking the Poisson
brackets $\{H,\varphi_i\}$ and $\{\varphi_i,\varphi_j\}$ they generate
secondary constraints. Let suppose that this process closes and at the
end we obtain a finite number of independent dynamical variables
($u_1,\dots,u_r)$ which describe the dynamics of the constrained
system. These dynamical variables satisfy a system of equations of the
following form
$$\{u_i,u_j\}=f_{ij}(u_1,\dots,u_r)\eqno(1.3)$$
$$i,j=1,2,\dots,r ,\qquad i<j$$
with $f_{ij}=-f_{ji} $  skew-symmetric functions. If $f_{ij}$ has a power
series expansion this may have the form
$$f_{ij}(u_1,\dots,u_r)= a_{ij}+b_{ij}^ku_k+c_{ij}^{kl}u_k u_l+\dots$$

In this approach we make no distinction between Hamiltonian, primary
and secondary constraints, first or second class constraints, all of
them are simply dynamical variables living in a democratic society,
the rules on which they obey being the system of equations (1.3). Now,
because the Hamiltonian is only one of the pairs, we have to solve the
problem of integrals of motion for a dynamical system governed by
Eqs. (1.3).
It seems natural to extend the classical solution, $F$ is an integral
of motion if $\{F,H\}=0$, to the new context by requiring that $F$
is an integral of motion if
$$\{F,u_i\}=0, \qquad i=1,2,\dots,r$$

We remind that the same condition was used by Dirac too \cite{Di},
{\it but only in the new symplectic structure, the Dirac bracket,
$\{\cdot,\cdot\}_D$, and not in the canonical Poisson structure as we do
here.}
 Taking into account the Poisson-Lie  structure defined by Eqs. (1.3)
the above equation is equivalent to the following system of first
order
partial differential equations
$$\sum_{i=1}^{i=r}{\partial F\over\partial
u_i}\{u_i,u_j\}=\sum_{i=1}^{i=r}{\partial F\over\partial
u_i}f_{ij}(u_1,\dots,u_r)=0\eqno(1.4)$$
$$ j=1,\dots,r$$

Being a homogeneous system a necessary condition for the existence of
a non-trivial solution, $F\not={\rm ct}$, is \cite{K,Go}
$$\det|\{u_i,u_j\}|=\det|f_{ij}(u_1,\dots,u_r)|=0\eqno(1.5)$$

The solution(s) of the system (1.4) will depend in general on all
dynamical variables and  will play the r\^ole of the classical
Hamiltonian for non-constrained systems, they being the conserved physical 
quantities of the
dynamical system. The classical theory of first order partial
differential equations tell us that if the rank of the system (1.4) is
$p$ then (1.4) may have up to $n=r-p$ independent solutions and the
easiest way to obtain them is by using the characteristic method
\cite{K,Go}. The simplest solutions of the system (1.4) are called
elementary solutions \cite{K,Go}, the general solution being an arbitrary
continuous and derivable function of these elementary solutions 
$G=G(F_1,F_2,\dots)$.
By quantisation $\{\cdot,\cdot\}$ goes into ${1\over
i\hbar}[\cdot,\cdot]$ and the observables of the theory will be the
solutions of the system (1.4). When the algebra (1.3) reduces to that of a
semi-simple Lie algebra the solutions $F_k$ will be the Casimir
operators of this algebra and if the respective algebra has rank $l$ 
there will be $l$
Casimir operators by the well-known result of Racah \cite{Ra}.
Thus Eqs. (1.3)-(1.4) represent a generalisation of the known powerful
machinery of representation theory of Lie algebras and give us a
method for finding the maximal set of commuting observables for a
given physical system. Finding the physically relevant operators and
their spectra is one of the goals of any quantum theory

\vskip3mm
2. In the following we shall illustrate the new method with a few
examples to show that its content is not void.
\vskip2mm
2.1 We consider first   the motion of a particle on a $n-1$-dimensional
sphere which is the toy model for testing quantum constrained dynamics
\cite{KS,PD,HH,AB}.
The free Hamiltonian is 
$$H={1\over 2}(p,p)= {p^2\over 2}$$
where $(p,p)$ denotes the Euclidean scalar product in the $n$-dimensional
space, i.e. $\sum_{i=1}^{i=n} p_i^2$ .
The primary constraint is usually written as
$$\varphi=(q,q)-R^2=r^2-R^2=0$$
The Eqs. (1.3) take the form

$$\{\varphi,H\}=2V\,\, \qquad
\{V,H\}=2H$$
$$\{ \varphi,V \}=2(\varphi+R^2)=2r^2 $$
where  $V=(q,p)$ is the secondary constraint. The system of differential 
equations is

$$-2V{\partial F\over\partial H}-2(\varphi+R^2){\partial
F\over\partial V}=0$$
$$2V{\partial F\over\partial\varphi}-2H{\partial F\over\partial V}=0\eqno(2.1)$$
$$2(\varphi+R^2){\partial F\over\partial\varphi}-2H{\partial
F\over\partial H}=0$$
The condition (1.5)  is satisfied the dimension of the matrix   being odd.
By applying the characteristic method \cite{K,Go} we have from the last 
equation
$$\varphi'(t)=\varphi+R^2\,\,\qquad H'(t)=-H$$
The solution  is 
 $$  \varphi+R^2=e^t  \,\, \qquad    H=e^{-t}$$
By eliminating  $t$ we find that the solution has the form
$$F=(\varphi+R^2)H+g(V)$$

If we use this form in the second equation we get $g(V)=-V^2/2$ .
Thus an elementary solution of the system (2.1) is
$$F=UH-V^2/2$$
where $U=\varphi+R^2=r^2=(q,q)$.

The "Hamiltonian" will be
$${\cal H}=UH-V^2/2={1\over 2}\left(\sum_1^nq_i^2\sum_1^np_i^2-
(\sum_1^np_iq_1)^2\right)=$$
$${1\over 2}\sum_{i<j}^n(q_ip_j-q_jp_i)^2={1\over
2}\sum_{i<j}^nL_{ij}^2={1\over 2}L^2$$
Thus the quantum observable is the square of the angular momentum 
\cite{PD,HH,AB}. 
Let show that ${\cal H}$ is the good classical Hamiltonian of the
problem.
The Hamilton equations
$$\dot{q_j}={\partial{\cal H}\over\partial p_j} \,\,\qquad
\dot{p_j}=-{\partial{\cal H}\over\partial q_j}$$
have the form
$$\dot{q_j}=p_j\,(q,q)-q_j\,(q,p)$$

$$\dot{p_j}=-q_j\,(q,q)+p_j\,(q,p)$$

Multiplying the first equation by $p_j$, the second by $q_j$ and
taking the sum we get
$$\dot{q_j}p_j+q_j\dot{p_j}={d\over dt}(q,p)={dV\over dt}=0$$
Similarly multiplying the first equation by $q_j$ we obtain
$$\dot{q_j}q_j={1\over 2}{dU\over dt}=(q,p)(q,q)-(q,q)(q,p)=0$$
which shows that $U$ and $V$ are constant in time and if the
constraints are fulfilled at the initial time they will be fulfilled
at any time. We consider the last two relations as a consistency check
of the formalism.
\vskip2mm
2.2 We consider now a more complicated structure, the functions
$f_{ij}$ entering Eqs. (1.4) being quadratic functions. One of  the
first
such a structure is that introduced by Sklyanin in connection with the
Yang-Baxter equations \cite{Sk}. The eqs. (1.4) have the following form
$$\{u_2,u_1\}=b_1 u_3u_4\quad\{u_3,u_1\}=b_2 u_2u_4$$
$$\{u_4,u_1\}=b_3u_2u_3\quad
\{u_3,u_2\}=a_1 u_1u_4$$
$$\{u_4,u_2\}=a_2 u_1u_3\quad\{u_4,u_3\}=a_3
u_1u_2$$
$a_i$ and $b_i$ , $i=1,2,3$ being arbitrary complex numbers. The case
considered by Sklyanin was $a_1=-a_2=a_3$ and $b_1+b_2+b_3=0$.
The system (1.5) has the form
$$b_1u_3u_4{\partial F\over\partial u_2}+b_2u_2u_4{\partial
F\over\partial u_3}+a_2u_1u_3{\partial F\over\partial u_4}=0$$
$$-b_1u_3u_4{\partial F\over\partial u_1}+a_3u_1u_4{\partial
F\over\partial u_3}+b_3u_2u_3{\partial F\over\partial u_4}=0\eqno(2.2)$$
$$-b_2u_2u_4{\partial F\over\partial u_1}-a_3u_1u_4{\partial
F\over\partial u_2}+a_1u_1u_2{\partial F\over\partial u_4}=0$$
$$-b_3u_2u_3{\partial F\over\partial u_1}-a_2u_1u_3{\partial
F\over\partial u_2}-a_1u_1u_2{\partial F\over\partial u_3}=0$$

The condition (1.5) is equivalent to
$$a_1b_1-a_2b_2+a_3b_3=0\eqno(2.3)$$
so in the following we suppose that (2.3) holds. From the first equation we 
have
$$u_2'(t)=b_1u_3u_4\,,\quad u_3'(t)=b_2u_2u_4\,,\quad u_4'(t)=b_3u_2u_3$$
From the first two relations we have that $b_1u_3^2-b_2u_2^2=ct$. This
suggest to look for a solution of the form
$$F(u_1,u_2,u_3,u_4)=b_1u_3^2-b_2u_2^2+g(u_1)$$
independent of $u_4$. The substitution of this $F$ in the second
equation (2.2) gives $g(u_1)=a_3u_1^2$ and the first Casimir has the form
$$C_1=a_3u_1^2-b_2u_2^2+b_1u_3^2$$
In the same manner one finds the second solution which is
$$C_2=a_1u_1^2-b_3u_3^2+b_2u_4^2$$
\vskip2mm
2.3 Another quadratic algebra is found in ref. \cite{BGST} used to
describe the kinematical symmetry of a spin chain on a one dimensional
lattice.It has the form
$$\{u_1,u_2\}=-{a\over 2}u_2^2\qquad\{u_1,u_3\}=a u_1\eqno(2.4)$$
$$\{u_2,u_3\}=a u_2\quad\{u_4,u_i\}=0\quad i=1,2,3$$
Since $u_4$ commutes with the other generators a solution of the
eqs. (1.4)) is of the form $f(u_4)$ with $f$ an arbitrary derivable function.
The other Casimir is
$$C=u_1 u_2^{-1}-{1\over 2}u_3^2$$

If we perturb the second equation (2.4) to the following form
$$\{u_1,u_3\}=a u_1+b u_2\eqno(2.2)$$
obtaining a Poisson-Lie structure on the 2-dimensional Galilei algebra
\cite{Ko},
the Casimir is more complicated and cannot be guessed simply. The
characteristic method gives
$$C=au_1u_2^{-1}-b\, log\, u_2-{a\over 2}u_3$$

\vskip2mm
2.4 An other interesting example appears in the construction of
Wess-Zumino-Witten models on non semi-simple groups \cite{NW}. The
algebra has the following structure
$$\{J,P_i\}=\epsilon_{ij}P_j\,,\quad \{P_i,P_j\}=\epsilon_{ij}T\,,$$
$$\{T,J\}= \{T,P_i\}=0\,, i=1,2$$
In general, given a Lie algebra to define a WZW model one needs a
bilinear form in the generators of the algebra, form  which is symmetric,
invariant and non-degenerate. Usually for semi-simple groups one takes 
Tr$\,u_iu_i$, with the trace taken in the adjoint representation of the
group. For nonsemisimple groups this quadratic form is degenerate. By
applying our formalism one finds easily the two Casimirs
$$C_1=P_1^2+P_2^2+2 JT\,,\quad C_2=g(T)$$
where $g(T)$ is an arbitrary derivable function of $T$.
Thus the most general bilinear form is
$$\Omega=a(P_1^2+P_2^2+2JT)+b T^2$$
where $a$ and $b$ are two arbitrary constants, which is the result of
Nappi and Witten.

\vskip2mm
2.5 Now we want to show that  finding the spectrum of the
hydrogen atom is also a problem of quantisation with constraints.
The classical Hamiltonian is 
$$H={p^2\over 2 m} -{\kappa\over r}$$
where $m$ is the reduced mass and $\kappa=Ze^2$. $H$ and the angular
momentum ${\bf L=r \times p}$ are constants of the motion. But these
quantities are not enough to make the orbit to be closed, and not
enough for having a discrete spectrum.   We
quote from Schiff's book \cite {Sc}

"The rotational symmetry of $H$ is enough to cause the orbit to lie in
some plane through $O$, but is not enough to require the orbit to be
closed. A small deviation of the potential energy from the Newtonian
form $V(r)=-({\kappa/ r})$ causes the major axis $PA$ of the
ellipse to precess slowly, so that the orbit is not closed. This suggests
that there is some quantity , other than $H$ and $L$, that is a
constant of the motion and that can be used to characterise the
orientation of the major axis in the orbital plane."

Such a quantity, which we see as a constraint, is the
Laplace-Runge-Lenz vector. It is proportional with the Di-polar
momentum of the orbit and has the form
$${\bf M}={{{\bf p\times L}}\over m}-{\kappa\over r}{\bf r}$$

These constraints generate the first quadratic algebra in quantum
physics. Indeed we have
$$\{M_i,M_j\}=-{2\over m}\epsilon_{ijk}HL_k\,\qquad i,j=1,2,3$$
where $H$ and $\bf L$ are the energy and, respectively, the angular
momentum.
The energy commutes with all the other quantities
$$\{H,M_i\}=\{H,L_i\}=0\, \qquad i=1,2,3$$
and we have also
$$\{M_i,L_j\}=\epsilon_{ijk}M_k\,,\quad \{L_i,L_j\}=\epsilon_{ijk}L_k\,,\quad
i=1,2,3$$

The Eqs. (1.4) have the form
$$-L_3{\partial F\over\partial L_2}+L_2{\partial F\over\partial L_3}-
M_3{\partial F\over\partial 
M_2}+
M_2{\partial F\over\partial M_3}=0$$

$$L_3{\partial F\over\partial L_1}-L_1{\partial F\over\partial L_3}+
M_3{\partial F\over\partial 
M_1}-
M_1{\partial F\over\partial M_3}-0$$
$$-L_2{\partial F\over\partial L_1}+L_1{\partial F\over\partial L_2}-
M_2{\partial F\over\partial 
M_1}+
M_1{\partial F\over\partial M_2}=0\eqno(2.5) $$
$$-M_3{\partial F\over\partial L_2}+M_2{\partial F\over\partial L_3}+
\alpha L_3{\partial F\over\partial 
M_2}-\alpha
L_2{\partial F\over\partial M_3}=0$$
$$M_3{\partial F\over\partial L_1}-M_1{\partial F\over\partial L_3}-
\alpha L_3{\partial F\over\partial 
M_1}+\alpha
L_1{\partial F\over\partial M_3}=0$$
$$-M_2{\partial F\over\partial L_1}+M_1{\partial F\over\partial L_2}+
\alpha L_2{\partial F\over\partial 
M_1}-\alpha
L_1{\partial F\over\partial M_2}=0$$
where $\alpha={2 H/ m}$. Since $H$ commutes with all the other
quantities it is in the centre of algebra and, such as in the previous
example,  it will be a Casimir, i.e. an observable in the quantum theory.
Thus $C_1=H=E$ is a good quantum number.

One can  easily see that $L^2$ and $M^2$ are separately solutions of
the first three equations (2.5), but none of them satisfies the last
three  equations. We look for a solution of the form
$$F=a\,L^2+b\,M^2$$
with $a$ and $b$ some constants. We find from the fourth equation that
$b/a=-1/\alpha$ and the second Casimir is
$$C_2=a(L^2-{M^2\over\alpha})$$
The third is 
$$C_3={\bf L\cdot M}$$
If we use the quantum form of $M^2$, i.e.
$$M^2={2E\over m}(L^2+\hbar^2)+\kappa^2$$
and take $a=1/4$ in the second Casimir, we find the known form of the energy 
levels
$$E=-{m \kappa^2\over 2 \hbar^2(2c+1)^2}$$
where $c=0,1/2,1,\dots$ are the  eigenvalues of $C_2$.
In conclusion the hydrogen atom {\it has} a symmetry group, but this
 is a nonsemisimple one, its Lie algebra has dimension seven,
and, more important, for explaining the discrete spectrum is not
necessary to invent a dynamical symmetry like $O(4)$.
\vskip3mm
3. The Dirac quantum theory \cite{Di} was patterned after the
classical theory, the "observables" representing constraints must have
zero expectation values. This requirement is not consistent with the
fact that the Poisson brackets between Hamiltonian and constraints and
between constraints themselves may not vanish such as 
Eqs. (1.3) show. In this paper we have shown that this inconsistency
disappears
if we postulate that the observables are the Casimir operators of the
algebra (1.3). A consequence of this postulate is the following,
starting
with a Dirac form Hamiltonian
$$H_D=H+\mu_i\varphi_i$$
may be misleading and cause troubles when using it for the description
of physical systems, the true Hamiltonians being more complicated
functions of both the old Hamiltonian and the constraints together, as
the above examples suggest. The lesson to be learnt is that for
constrained systems almost no one of the initial dynamical variables
transforms into an observable. In this respect the hydrogen atom is an
exception, the reason being that the classical Hamiltonian commutes
with all the constraints, being in the centre of the Poisson-Lie group.
\vskip3mm
Acknowledgements. This work was done while the author was a visitor at
Institute of Theoretical Physics, University of Bern in the frame of
the Swiss National Science Foundation program "Cooperation in Science
and  Research with Central and Eastern European Countries and New
Independent States 1996-1998. Institutional Partnership". I take this
opportunity to thank the Swiss  National Science Foundation for
support. The warm hospitality of Professor H. Leutwyler is gratefully 
acknowledged.

\end{document}